# A Techno-Economic Analysis of Asteroid Mining

**Andreas M. Hein**[a]*, **Robert Matheson**[b], **Dan Fries**[c]

[a] *Initiative for Interstellar Studies, Bone Mill, Charfield, United Kingdom, 31241*, Andreas.hein@i4is.org
[b] *Initiative for Interstellar Studies, Bone Mill, Charfield, United Kingdom, 31241*, robmatheson95@gmail.com
[c] *Initiative for Interstellar Studies, Bone Mill, Charfield, United Kingdom, 31241*, dan_fries@web.de
* Corresponding Author

**Abstract**

Asteroid mining has been proposed as an approach to complement Earth-based supplies of rare earth metals and supplying resources in space, such as water. However, existing studies on the economic viability of asteroid mining have remained rather simplistic and do not provide much guidance on which technological improvements would be needed for increasing its economic viability. This paper develops a techno-economic analysis of asteroid mining with the objective of providing recommendations for future technology development and performance improvements. Both, in-space resource provision such as water and return of platinum to Earth are considered. Starting from first principles of techno-economic analysis, gradually additional economic and technological factors are added to the analysis model. Applied to mining missions involving spacecraft reuse, learning curve effect, and multiple spacecraft, their economic viability is assessed. A sensitivity analysis with respect to throughput rate, spacecraft mass, and resource price is performed. Furthermore, a sample asteroid volatile mining architecture based on small CubeSat-class spacecraft is presented. It is concluded that key technological drivers for asteroid mining missions are throughput rate, number of spacecraft per mission, and the rate in which successive missions are conducted.

## 1. Introduction

The exploitation of asteroids, and in particular Near Earth Asteroids (NEAs) has been repeatedly proposed as a source of resources for Earth and space [1]–[3]. Ross [4] distinguishes between metals and volatiles as resources along with their use in a variety of applications such as construction, life support systems, and propellant. In particular, volatiles have received attention for in-space use, due to their relative ease of extraction. For example, Calla et al. [5] explored the technological and economic viability of supplying water from NEAs to cis-lunar orbit.

Regarding the supply of resources to Earth, only resources with a high value to mass ratio are interesting, due to the high cost of returning such material. Therefore, high-value metals such as rare earth metals and in particular the subgroup of platinum group metals have been the subject of mining studies [6]. The supply of rare earth metals is crucial for many "green technologies" such as fuel cells, catalyzers, high-capacity batteries, and solar cells [7]–[9].

Existing research on asteroid mining has mainly looked into its economic viability [2], [6], [10], [11], technological feasibility [2], [12]–[17], cartography of asteroids [18], [19], and legal aspects [20]–[22]. More recently, environmental arguments for asteroid mining have been made, in particular with regards to platinum group metals [23]–[25].

In the following, we will focus on the economic viability of asteroid mining. Andrews et al. [6] have analysed the economic viability of a specific asteroid mining architecture, using Net Present Value (NPV) analysis. Busch [10] provides an analysis of an asteroid mining architecture using magnetic sails and high-efficiency solar cells. Sonter [2] provides a generic cost analysis framework for asteroid mining, linking economic parameters to technical parameters. Figures of merit for comparing asteroid mining architectures are discussed in Lewis [26], [27]. A key figure of merit is the Mass Payback ratio (MPBR) [28]. Oxnevad [29] proposes NPV for going beyond MPBR by including development cost and cost of capital. Diot and Prohom provide an overview of various strategic, financial, and supply chain implications for asteroid mining [30]. Gertsch and Gertsch [31] use return on investment (ROI) analysis to a sample asteroid mining venture. Craig et al. [32] provide a year-by-year cash flow analysis for a mining mission, using the M-type NEA 1986 DA as a case study. They conclude that a mining venture would currently be too risky to commercially succeed.

It seems that the NPV model developed in Sonter [2] is currently the most detailed model which links economic to technical parameters. Nevertheless, the model has shortcomings:

- Only a single mining mission is considered;
- Cost of return to Earth surface is omitted;
- Development cost is not considered;
- Focused on impulsive transfers.

In this article, we will provide a techno-economic analysis of asteroid mining in order to identify key criteria for economic viability. For this purpose, we specifically focus on economic viability compared to immediate alternatives, such as terrestrial mining and direct delivery of resources from Earth to space. For this purpose, we derive a generic profitability model for analysing the economic viability of asteroid mining and apply it to the case of water and platinum mining. In addition, we analyse supply-demand effects that could follow from the injection of platinum into the terrestrial market, taking price elasticity and substitution into account.

**2. Material and methods**

We modify and extend the techno-economic model from Sonter [2], first looking at the profit from a single mission, ignoring the time value of money. Subsequently, a profit equation with spacecraft reuse is proposed. Next, a profit equation taking multiple spacecraft into consideration is developed, including reuse. We further add a learning curve for the production cost of spacecraft. Subsequently, the time-value of money is taken into account. The resulting profit equationl addresses several of the shortcomings of Sonter's [2] model mentioned in the introduction and is capable of :

- Assessing multiple mining missions as part of an entire asteroid mining venture;
- Taking the cost of return to Earth into account;
- Including development cost;
- Taking transfer time into account via a generic mission start date and the date for resource return.

The resulting profit equation is applied to the case of water and platinum mining.

For the supply-demand analysis for platinum, a linear supply-demand curve is developed, including price elasticity. In addition, an equation for the total amount of platinum in the global terrestrial market is defined, which is used for expressing the substitution of terrestrial platinum by platinum from space. Further, using a simplified version of a profit equation, the profitability of a platinum mining venture is assessed.

**3. Theory and calculation**

*3.1 Asteroid mining breakeven analysis*

We start with a simple equation where profit $P$ from a single mission is revenue $R$ minus cost $C$. We neglect all upfront costs such as development costs and limit $R$ to revenue from material sold, writing $R_{mat\_sold}$. As shown in equation 1, we can further expand $R_{mat\_sold}$ to the mass of material $m$, multiplied by the mass-specific price $c_{price\_mat}$ of that material. We further break down $C$ into production cost $C_{prod}$, transportation cost from Earth into space and in space $C_{transport}$, and cost of operations $C_{ops}$.

$$P = R_{mat\_sold} - C =$$
$$m_{mat} c_{price\_mat} - (C_{prod} + C_{transport} + C_{ops}) \quad (1)$$

The expression in equation 1 is sufficiently general that it can also be used for the case where resources are brought into space from Earth. In such a case, it can be seen that the price of the resource sold needs to at least cover the incurred cost. The equation is equally valid for the case where material is mined in space and brought back to Earth. In such a case, the price of a resource would be the market price on Earth.

We express production and transportation cost in terms of the spacecraft mass $m_{SC}$ and the mass-specific production and launch cost $c_{prod}$ and $c_{transport}$ respectively. Note that $c_{transport}$ only takes the cost of transport up to the point into consideration, where external transportation services are used.

$$P = m_{mat} c_{price\_mat} - m_{SC} * (c_{prod} + c_{transport}) - C_{ops} \quad (2)$$

The mass of material $m_{mat}$ is expressed in terms of the spacecraft dry mass $m_{SC}$, the throughput rate $f$ in kg processed per second per kg of spacecraft mass, time of mining t, and the fraction of extracted resources from the total material processed $r$.

$$m_{mat} = m_{SC} * f * t * r \quad (3)$$

Contrary to distinguishing between the mass of the processing equipment of the spacecraft in Sonter [2], we treat the overall dry mass of the spacecraft as a reference, as the dry mass is less susceptible to change during a mission than the wet mass. Note that the choice of reference is arbitrary and we could similarly choose the initial wet mass of the spacecraft as a reference. In that case, the value for $f$ would change, with respect to the reference. Inserting equation 3 into 2 and rearranging gives:

$$P = m_{SC} * (f * t * r * c_{price\_mat} - (c_{prod} + c_{transport})) - C_{ops} \quad (4)$$

Note that in case the spacecraft dry and wet mass are different, $c_{transport}$ has to be adjusted by a mass ratio R, indicating the ratio between spacecraft wet mass and dry mass. In particular, for rocket propulsion systems, the mass ratio is determined by the Tsiolkovsky equation. Rearranging (4) yields:

$$P = m_{SC} * c_{price\_mat} * [f*t*r - (\frac{c_{prod}+c_{transport}}{c_{price\_mat}})] - C_{ops} \quad (5)$$

As a minimum condition, the expression in [] must be > 0, $f*t*r$ needs to be sufficiently high, and in addition, the expression in () as small as possible.

Equations 5 is important, as it defines criteria for minimum economic viability. If the bracket [] is not positive, no profit is generated. We can also see that we could, in principle, increase the mining duration $t$ arbitrarily and could always yield a profit for positive $f$ and $r$, no matter the associated cost. This is obviously not realistic and realistic values for $t$ are likely to be less than 5 years. Another observation is that there are essentially two key technical parameter that are under the direct influence of the asteroid mission architect. The first is the throughput rate $f$, which should be as high as possible, and second, $c_{prod}$, which should be as low as possible. Note that the duration of transportation has been neglected in the equation.

Up to this point, only operational costs, also called operational expenditures (OpEx) were considered, as we have focused on a single mission. By a single mission, we understand a single spacecraft performing the mining operation and transporting the refined material to the location where it is transferred to a client.

However, in a realistic scenario, the capital costs (or capital expenditures CapEx) such as the development cost of the mining spacecraft would also need to be taken into account. The most dominant contribution to CapEx is likely to be the development cost of the spacecraft. We then yield for the total profit $P_{tot}$ for the mining venture:

$$P_{tot} = n*[m_{SC}*c_{price\_mat}*[f*t*r - (\frac{c_{prod}+c_{transport}}{c_{price\_mat}})] - C_{ops}] - C_{dev} \quad (6)$$

The factor $n$ indicates the number of mining missions, under the assumption that one spacecraft is used for one mission. $C_{dev}$ indicates the development cost. In the following, we will express development cost $C_{dev}$ in the form of $m_{SC}*c_{dev}$, where $c_{dev}$ is the mass-specific development cost. For the case where spacecraft are reused $s$-times, we yield:

$$P_{tot} = n*\left[m_{SC}*c_{price_{mat}}*\left[\left(f*t*r - ceil(\frac{n}{s+1})\left(\frac{c_{prod}+c_{transport}}{c_{price_{mat}}}\right)\right)\right] - C_{ops}\right] - m_{SC}*c_{dev} \quad (7)$$

Where $ceil()$ is a ceiling function, which selects the smallest following integer. For the missions where the spacecraft is reused, $c_{prod}$ and $c_{transport}$ disappear. Note that this equation does not contain the duration of the venture.

We can introduce time in terms of years by summing up $n_i$ and $s_i$ per year, counting the number of missions delivering resources in a year $i$ and analogously for the number of missions with a reused spacecraft $s_i$.

$$P_{tot} = \sum_{i=1}^{T} n_i * \left[m_{SC}*c_{price_{mat}}*\left[\left(f*t*r - ceil(\frac{n_i}{s_i+1})\left(\frac{c_{prod}+c_{transport}}{c_{price_{mat}}}\right)\right)\right] - C_{ops}\right] - m_{SC}*c_{dev} \quad (8)$$

Breakeven can be calculated by putting $P_{tot} = 0$ and solving the resulting implicit function for $n$. Note that an underlying assumption of equation (8) is that the $c_{prod}$ and $c_{transport}$ are incurred in the same year as the mined resources are sold on the market. Otherwise, these cost factors would need to be accounted for in a different year. As we are assuming here that there is no time value of money, it does not make a difference with regards to $P_{tot}$. We will consider the more general case with the time value of money in Section 3.2.

Going one step further, manufacturing the same or similar types of spacecraft multiple times leads to the reduction of production cost, also called experience curve or learning curve. We can introduce the experience curve by a typical power law relation.

$$c_{prod\_n} = c_{prod\_1} * n^a \quad (9)$$

Where $c_{prod\_n}$ is the specific production cost of the nth unit, $c_{prod\_1}$ the specific production cost of unit 1, $n$ the number of units produced and $a$ the learning factor, defined as:

$$a = \frac{\log(p)}{\log(2)} \quad (10)$$

With $p$ typically 0.85 for aerospace systems, we get a = -0.234. The resulting equation is:

$$P_{tot} = \sum_{i=1}^{T} n_i * \left[ m_{SC} * c_{price_{mat}} * \left[ \left( f*t*r - ceil(\frac{n_i}{s_i+1}) \left( \frac{c_{prod\_1}*[ceil(\frac{n_i}{s_i+1})]^a + c_{transport}}{c_{price_{mat}}} \right) \right) \right] - C_{ops} \right] - m_{SC} * c_{dev} \quad (11)$$

For the case where multiple spacecraft per mission are used, we introduce a factor *p* that indicates the number of spacecraft per mission, assuming spacecraft of the same type. In the case of multiple spacecraft per mission, $m_{SC}$ is the total mass of all spacecraft participating in a mission. Regarding the learning curve, we assume that all spacecraft for a mission are developed in one batch, therefore benefit from the same learning curve factor, and are of the same type. The resulting equation is:

$$P_{tot} = \sum_{i=1}^{T} n_i * \left[ m_{SC} * c_{price_{mat}} * \left[ \left( f*t*r - ceil\left(\frac{n_i}{s_i+1}\right) \left( \frac{c_{prod_1}*\left[p*ceil\left(\frac{n_i}{s_i+1}\right)\right]^a + c_{transport}}{c_{price_{mat}}} \right) \right) \right] - C_{ops} \right] - c_{dev} * \frac{m_{SC}}{p} \quad (12)$$

From equation (12), it can be seen that increasing the number of spacecraft per mission tends to decrease production cost due to the learning curve effect and decreases development cost. We assume that advanced automation enables to keep operations cost constant, even with an increased number of spacecraft per mission. In case this assumption is not valid, $C_{ops}$ needs to be multiplied by a function depending on *p*.

*3.2 Asteroid mining net present value analysis*
In the breakeven analysis in 3.1, we have omitted the time value of money for economic viability. For a more realistic analysis, we introduce the present value of money via $(1 + I)^i$, where *I* is the interest rate and *i* the year.
Similar to Sonter [2], the present value of a revenue stream from an asteroid mining mission can be calculated via:

$$P_{PV} = \frac{R}{(1+I)^t} - C \quad (13)$$

*I* is the interest rate, *t* the year of the revenue stream and *C* the cost incurred in year 0. The equation is only valid for a single revenue stream in year *t* and only considers incurred cost at the beginning of the project. We can remedy for these shortcomings by first extending the equation to multiple missions.

$$P_{PV} = \sum_j \frac{R_j}{(1+I)^{t_j}} - C_j \quad (14)$$

Where j indicates the j-th mission.
Second, the equation is further extended to consider missions that do incur cost at different points in time, for example production and launch cost of the spacecraft. We therefore add $m_j$, which is subtracted from the revenue stream $t_j$. $m_j$ is usually the mission duration. This leads to:

$$P_{PV} = \sum_j \frac{R_j}{(1+I)^{t_j}} - \frac{C_j}{(1+I)^{t_j-m_j}} \quad (15)$$

We insert the respective revenue and cost factors from (12) into (.) and get for the present value for the total profit over all missions $P_{tot\_PV}$:

$$P_{tot\_PV} = \sum_{i=1}^{T} n_i \sum_j * \left[ m_{SC} * c_{price_{mat}} * \left[ \left( \frac{f*t*r}{(1+I)^i} - ceil\left(\frac{n_i}{s_i+1}\right) \left( \frac{c_{prod_1}*\left[p*ceil\left(\frac{n_i}{s_i+1}\right)\right]^a + c_{transport}}{c_{price_{mat}}*(1+I)^{i-m_j}} \right) \right) \right] - C_{ops} \right] - c_{dev} * \frac{m_{SC}}{p} \quad (16)$$

From exploring these basic relationships, we can find several technical parameters we can tune for maximizing profit:
- Use multiple spacecraft per mission for exploiting learning curve effects, which result in lower development cost;
- Increase the throughput rate;
- Reduce mission duration;
- Compress the sequence of missions (launch follow-up spacecraft before previous spacecraft returns);
- Increase reuse of spacecraft
- Reduce other cost factors such as operations and transportation cost.

Obviously, the magnitude of these factors depends on the concrete parameter values that are plugged in. Sample values will be given in the Results section . However, reducing development cost and increasing throughput rate seem to be the most promising approaches for increasing profitability.

*3.3 Supply and demand-dependent profitability*

In the following, we develop the framework for analysing the profitability of an asteroid mining venture with respect to supply and demand. The first assumption in the supply-demand model is that there is only one firm that is mining asteroids for platinum, rather than multiple competing firms.

In the model I estimate the demand curve for platinum is estimated as linear with the equation:

$$C_{price\_mat} = \alpha + \beta(M_{tot}) \quad (11)$$

Where $C_{price\_mat}$ is the price of platinum in the global market as covered earlier, $\alpha$ is a constant, the intercept of the curve on the price axis, $\beta$ is a constant and the slope of the curve which is less than zero as demand curves slope downward and $M_{tot}$ is the total quantity of platinum supplied to the world markets in one year, in kg.

We can approximate the platinum demand curve to being linear as we are trying to capture how sensitive profitability is to the price elasticity of platinum in the local vicinity of the global platinum demand curve, where elasticity is the negative of the percentage change in price that occurs with a percentage change in quantity. The slope of the demand curve is directly proportional to the elasticity at any point on the demand curve, so $\beta$ is the first lever in my model.

To practically estimate this demand curve, five-year averages of the price of platinum per kg are used [33] and global demand for platinum [34] to estimate the location of a single point on the global platinum demand curve ($C_{mat_1}$, $M_{tot_1}$) = ($40,449 per kg, 254,582kg). It is known that platinum, like most rare earths has an inelastic price elasticity of demand (PED), this restricts the range of elasticities to 0<PED<1, to stay within this range we restrict the range of $\beta$ to -0.5> $\beta$ >-0.6. Given we know $\beta$ and that all demand curves go through point ($C_{mat_1}$, $M_{tot_1}$), we can derive corresponding values of $\alpha$ for each $\beta$.

The equation for $M_{tot}$ is:

$$M_{tot} = M_{earth} + n * M_{mat} \quad (12)$$

Where $M_{earth}$ is the mass of platinum supplied into the global markets in a year from sources on Earth, from mining and recycling, *n* is the number of platinum asteroid mining missions run in one year by our asteroid mining company, and $M_{mat}$ as stated earlier is the mass of platinum returned to earth from one mission. However, $M_{earth}$ is dependent on $n * M_{mat}$, we estimate this in the following manner:

$$M_{earth} = M_{tot_1} - k * n * M_{mat} \quad (13)$$

$M_{tot_1} = 254,582 kg$ is the quantity of platinum that would be supplied by earth-based platinum producers to the global market if the asteroid miner did not enter the market. $k$ is the mass of platinum by which earth production decreases if asteroid mining production increases by 1 kg, for example were $k = 0.5$ then every kg of platinum the asteroid miner produced would result in earth-based production of platinum decreasing by 0.5 kg. The production decisions of earth-based platinum producers depend on the quantity of platinum supplied by the asteroid miner into the global market. If $nM_{mat}$ increases, the total quantity of platinum in the global market increases and so the price of platinum decreases, this decreasing price will cause less profitable earth-based platinum mining producers to cease or reduce platinum production once again increasing the price.

Here an assumption is made in the model that earth-based platinum producers believe that the asteroid miner will continue to produce in the long-term and not go bankrupt. Given the long time delay between investment in platinum extraction and the eventual return on investment it is reasonable to believe that earth-based platinum producers will not change their production behaviour in response to the asteroid miners production level unless they believe the asteroid miner will continue to operate in the market long run.

This assumption is reasonable since asteroid mining is such a large feat it would likely only be successfully brought to fruition by a large and profitable company with enough funds for the R&D spend and investing in such a long-term project, and due to the political implications of rare earths the venture may also be backed by a government.

$k$ is the second lever in the model, it determines how platinum production of asteroid miners affects the quantity of platinum in the global markets, therefore the price of platinum and therefore the profitability of the asteroid mining venture. The closer earth-based platinum producers are operating to their profit margins the larger $k$ will be. A reasonable range for $k$ is 0<k<1.

The demand curve faced by the asteroid miner is presented below:

$$C_{price\_mat} = \alpha - \beta(M_{tot_1} - k * n * M_{mat} + n * M_{mat}) \quad (14)$$

This is inputted into the below profit function for the asteroid miner:

$$P(n) = n\big(M_{mat} * C_{price\_mat} - M_{SC}(C_{prod} + C_{transport}) - C_{ops}\big) - 0.1 C_{dev} = max! \quad (15)$$

Where the asteroid miner chooses the number of missions to launch per year, *n*, to achieve their maximum profit per year, *P*. We modelled the asteroid miner as having borrowed the funds necessary to finance the cost of development, $C_{dev}$. As a result the asteroid miner pays 10% fixed rate interest on the loan per year, $-0.1 C_{dev}$, which is generously low given how risky an asteroid mining venture would be. The asteroid miner does not pay off any of the principle of the loan, which is a reasonable assumption since the asteroid miner might like to keep its profits in a liquid form in case unexpected costs occurred such as a mission failure.

$\beta$ and $k$ are the only factors extrinsic to the model, the levers of our model. All other figures are either generated within the model, such as $\alpha$ and $n$, or are inputted constants from either the conservative costs column of Table 2 in this paper or from platinum market data from the World Bank and the World Platinum Investment Council.

In this model it is assumed that the demand curve is constant and does not shift to the left or right. However, there may be a feedback loop between supply and demand. If agents in the economy judge that due to asteroid mining the supply of platinum would be able to keep pace with growth in demand, and ensure the price remains relatively stable rather than rising wildly as would happen if demand grew whilst supply stayed fixed, then they might start investing in capabilities that require platinum. For example, petrol car manufacturers might consider using platinum rather than palladium in their catalytic converters. However, the timescales over which this might happen are large enough that other factors will likely be of greater significance, such as electric cars becoming more prominent and taking market share from the traditional automobile industry which in the period 2013-2017 constituted 41% of global platinum demand [34].

One limitation of this analysis is that platinum recycling has not been taken into account.

## 4. Results

### 4.1 Profitability for scenarios

For providing sample values for asteroid mining profitability, we use Equation (12) and omit interest rate. We use a conservative and optimistic scenario for water mining and platinum mining. The conservative scenario uses current cost values and the optimistic scenario uses cost values for mass production, similar to those which have been observed for the OneWeb constellation spacecraft. We assume a propellant-less propulsion system such as solar sails or electric sails, which have been proposed for asteroid exploration missions. As mentioned in Section 4.1, the use of a rocket propulsion system would require the introduction of a mass ratio, which would impact transportation cost and bootstrapping factor, which we will introduce in the next paragraph. Note that the reference for calculating the throughput rate, development cost is the dry mass of the spacecraft.

For water mining in Table 1, a value of 5% for the fraction of extracted resources versus processed material is used, which is a rather conservative value for C-type asteroids. The mining duration is set to 100 days. A small spacecraft with a mass of 150 kg is assumed, consistent with a recent proposal of using small spacecraft for asteroid mining [5]. The throughput rate for arriving at a reasonable duration to breakeven is $2.3 \times 10^{-4}$ kg/s/kg, meaning that for each kg of spacecraft dry mass, 0.23 grams of material are processed per second. Intuitively, this value does not seem to be excessively high. We also introduce a bootstrapping factor, where the total mass of water or platinum delivered is divided by spacecraft mass.

Table 1: Water mining conservative and optimistic scenarios

|  | Conservative cost (SMAD, Calla et al., 2017) | Optimistic cost (mass production) |
|---|---|---|
| Specific production cost [$/kg] | 1000000 | 10000 |
| Specific transportation cost to / from asteroid [$/kg] | 35000 | 35000 |
| Price of water [$/kg] | 20000 | 20000 |
| Operation cost [$] | 5700000 | 100000 |
| Specific development cost [$/kg] | 5453333 | 500000 |
| Mining duration [s] (100 days) | 8640000 | 8640000 |
| Ratio refined resources vs processed material | 0.05 | 0.05 |
| Throuput rate [kg/s/kg equipment] | $2.3 \times 10^{-4}$ | $2.3 \times 10^{-4}$ |
| Spacecraft mass [kg] | 150 | 150 |
| **Profit [M$]** | **139** | **293** |
| Material processed [kg] | 42944 | 1868 |
| Bootstrapping factor | 14.3 | 0.62 |
| **Years to breakeven [y]** | **5.9** | **0.26** |

We further add different assumptions to the reference case: 2-times reuse of spacecraft, learning curve effects, and multiple spacecraft per mission. The results for the conservative scenario can be seen in Figure 1. Only the number of missions is shown on the x-axis and mission duration is not indicated. However, as mining duration is fixed to 100 days and getting to and back from the asteroid(s) is likely to take several hundred days, mission duration can be assumed to be about one year. Furthermore, if the same spacecraft is reused, the spacecraft has to make a full cycle (outbound and inbound) before delivering resources. For the case of expendable spacecraft, launching spacecraft in tighter sequences could be imagined (launching subsequent spacecraft before return of precedent), compressing the sequence of missions. For the case of using multiple spacecraft, we assume 10 spacecraft per mission with a mass of 15 kg each.

Note that the profit for cases with reference, reuse, and learning all start with the same negative profit, as the development cost for all these cases is equal. However, using multiple, lighter spacecraft results in a lower development cost and breakeven is reached earlier. Learning effects seem to have a rather minor effect on profitability, as the comparison of the data points for two-times reuse and two-times reuse plus learning curve effect cases reveal.

Two sensitivity analyses are conducted, where the conservative reference case is varied by reducing the throughput rate by 50% or reducing the spacecraft mass by 50%. It can be seen that the reduction of the throughput rate has a large effect on profitability and would it render not profitable.

Another observation is that if spacecraft are expendable, missions could be launched in much tighter sequence than in the case of reuse, where the spacecraft has to make a full cycle of getting to and back from the asteroid. One can see that with launching missions twice as often as for the case of reuse, expendable spacecraft would be competitive with the reuse and reuse plus learning curve effect cases.

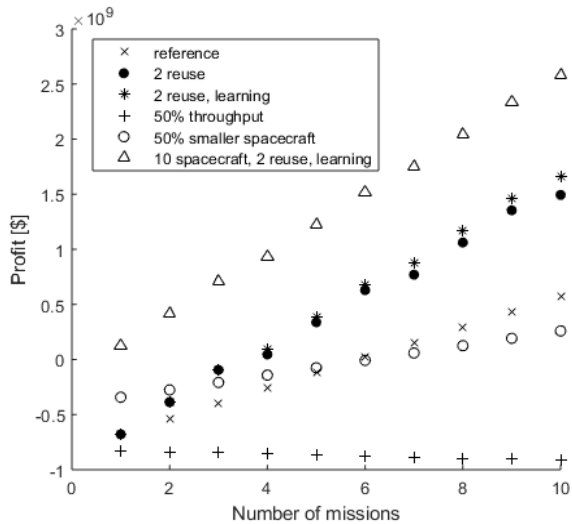

Figure 1: Water asteroid mining conservative scenario

A similar result is obtained for the optimistic scenario in Figure 2, where the reference case and those with reuse, learning curve effects, and multiple spacecraft are very close to each other. A similar observation can be made for the 50% reduction cases. The reason is the low cost, which makes the cases only dependent on throughput rate, price of resource sold, and spacecraft mass. An interesting observation is that for drastically lower costs, compressing the launch schedule for expendable spacecraft would lead to significant gains in profit. Hence, using expendable spacecraft would be better than reusing spacecraft for the optimistic scenario.

For platinum mining, the situation is quite different, simply due to the much lower ratio of platinum to asteroid material. We use a value of $10^{-5}$ from Blair [35]. For a similar breakeven to the water mining case, the throughput rate needs to be about two orders of magnitude higher. The value of 0.35 kg/s/kg seems very high. The spacecraft would need to be able to process the equivalent of its own mass within 3 seconds. For the optimistic scenario, this value could be reduced by one order of magnitude in case a similar breakeven as for the conservative scenario is aimed at.

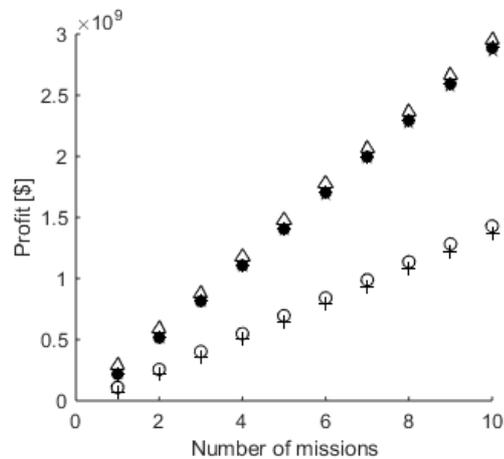

Figure 2: Water asteroid mining optimistic scenario

The platinum mining case does take the cost of returning the material to Earth into account via multiplying the launch cost by a factor two, which is a rough estimate of transporting platinum to the Earth surface. To include the cost of returning the material to Earth, the spacecraft mass needs to be increased by the mass used for e.g. aerobreaking and Earth re-entry.

Table 2: Platinum mining conservative and optimistic scenarios

|  | Conservative cost (SMAD, Calla et al., 2017) | Optimistic cost (mass production) |
|---|---|---|
| Specific production cost [$/kg] | 1000000 | 10000 |
| Specific transportation cost to / from asteroid [$/kg] | 35000 | 35000 |
| Price of platinum [$/kg] | 70000 | 70000 |
| Operation cost [$] | 5700000 | 100000 |
| Specific development cost [$/kg] | 5453333 | 500000 |
| Mining duration [s] (100 days) | 8640000 | 8640000 |

| Ratio refined resources vs processed material | $10^{-5}$ | $10^{-5}$ |
|---|---|---|
| Throuput rate [kg/s/kg equipment] | 0.35 | 0.35 |
| Spacecraft mass [kg] | 150 | 150 |
| **Profit [M$]** | **149** | **303** |
| Material processed [kg] | $6*10^7$ | $2.7*10^6$ |
| Bootstrapping factor | 4 | 0.2 |
| **Years to breakeven [y]** | **5.5** | **0.25** |

As for the case of water, we further add different assumptions to the reference case: 2-times reuse of spacecraft, learning curve effects, and multiple spacecraft per mission. The results for the conservative scenario can be seen in Figure 3 and Figure 4. Similar results as for the water case are yielded, as the throughput rate has been adjusted to the lower concentration of platinum. The transportation cost differs from the water case, as the platinum is transported to Earth.

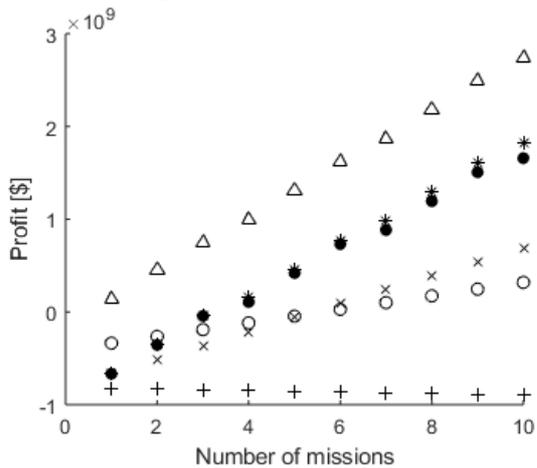

Figure 3: Platinum mining conservative case

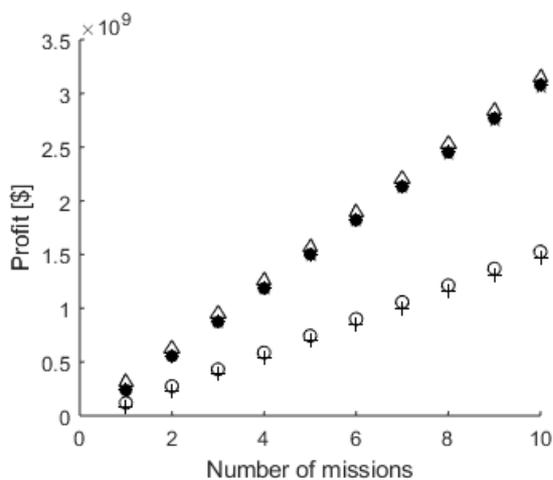

Figure 4: Platinum mining optimistic case

As fluctuations to platinum price are expected, we generate results for the case where platinum price stays at levels of $30,000 per kg instead of $70,000. These results can be seen in Figure 5 and Figure 6. For the conservative scenario in Figure 5, only the case with 10 small spacecraft per mission reaches breakeven within 10 years. For the optimistic case, the impact is limited and breakeven is reached for all cases in the first year, except the case with 50% throughput rate. It can be therefore seen that long-term prices of platinum would have a significant impact on asteroid platinum mining viability.

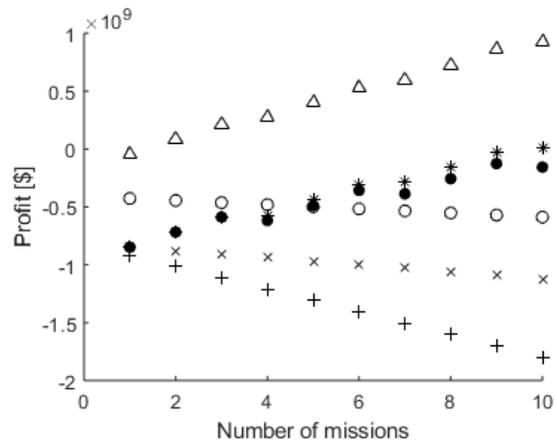

Figure 5: Platinum mining conservative case with $30k platinum price

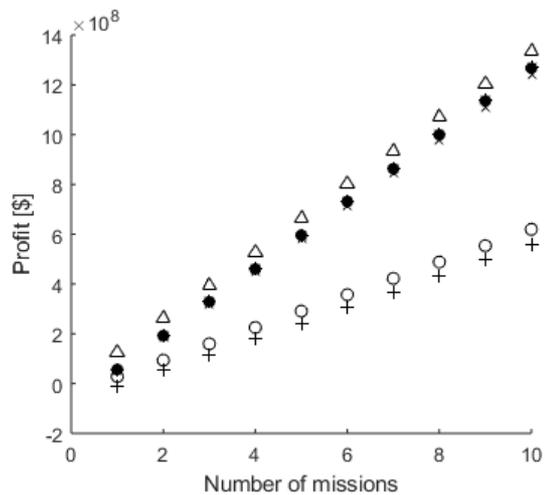

Figure 6: Platinum mining optimistic case with $30k platinum price

To summarize, for all analysed cases, using multiple small spacecraft combined with reuse and learning curve effect outperforms its alternatives. However, similar performance levels could be reached by the reference case for the optimistic scenario by compressing the launch schedule for expendable spacecraft by e.g. twice

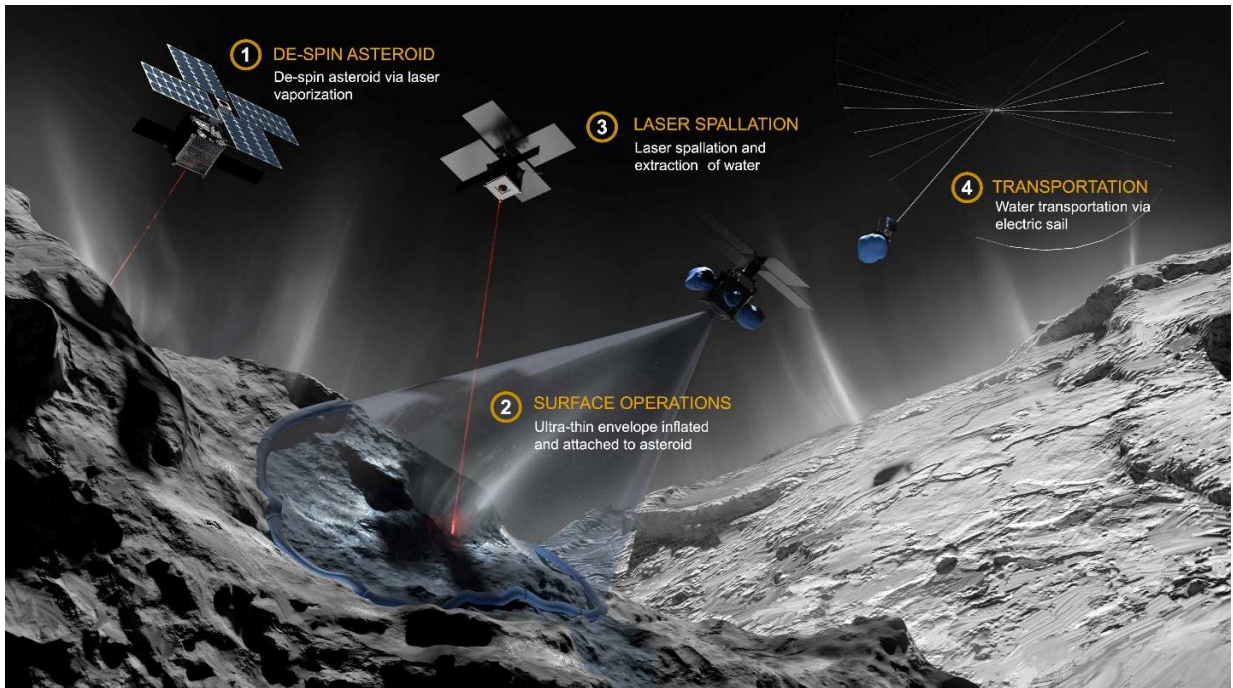

Figure 7: Small spacecraft asteroid volatile mining architecture using lasers for extraction and electric sails for transportation (Image credit: Efflam Mercier)

the rate of its alternatives. Overall, the technical parameter with the largest impact on economic viability is the throughput rate, followed by the number of spacecraft used per mission, which has a significant impact on development cost.

A sample mission architecture using multiple small spacecraft is shown in Figure 7. The spacecraft are expected to be 26 unit CubeSats. Two types of spacecraft are used: A laser spacecraft that heats the regolith with a laser and a volatile capturing spacecraft that captures and condenses the volatiles and transports them to their target destination using propellant-less propulsion such as an electric sail or solar sail [36]–[42]. As shown in Figure 7, the asteroid is first de-spinned (step 1) [43]–[47], then mined (step 2 and 3) and the mined resources transported to the target destination (step 4). In this architecture, lasers are used for extracting volatiles from the asteroid regolith and captured by transparent funnels with cold traps attached at their ends. Transportation is carried out via electric sails that do not consume any propellant.

*4.1 Supply-demand dependent profitability*

The table below shows the outputs of the supply-demand model based on its two input:

$\beta$ – directly proportional to the price elasticity of demand for platinum and determines how the total supply of platinum sold in the market per year affects the price of platinum. Therefore, it influences how the quantity of platinum produced by the asteroid miner affects the price it receives for it, and hence the profit of the asteroid miner.

$k$ – The decrease in earth-based platinum production per kg increase in asteroid mining platinum production. $k$ determines how the asteroid miner's platinum production affects the total quantity of platinum sold on world markets, which affects the price of platinum, which affects the asteroid miner's profitability.

Note: Breakeven is defined as the number of years of operation necessary for NPV = 0 to be true, with a discount rate of 10%.

It can be seen from the table that there is only a slim range of $\beta$ and $k$ values for which asteroid mining is profitable, and an even slimmer range in which it breaks even before 30 years of operation.

Where k = 1 the asteroid miner is profitable and breaks even in a year regardless of the value of $\beta$. However, this is an unlikely corner solution. There is a good probability that k is not greater than 0.9, it may be in the region of 0.6 or 0.4. Moreover, there is a good probability that the elasticity is not lower than 0.25.

This model has assumed that the quantity of platinum in the global market is the only factor affecting its price. However, platinum is a speculative asset as well as a useful rare earth and precious metal. Its price is subject not only to supply and demand, but to speculation. For example, the prices of precious metals are sensitive to market expectations of global, especially American, interest rates. The higher interest rates are the more

Table 3: Profitability analysis of asteroid mining, taking terrestrial production decrease into account.

| Elasticity | Slope of demand curve | \multicolumn{14}{c|}{Decrease in Earth based platinum production per kg increase in asteroid mining platinum production [kg]} |
| --- | --- | --- | --- | --- | --- | --- | --- | --- | --- | --- | --- | --- | --- | --- | --- |
| | | 1 | 0.99 | 0.98 | 0.97 | 0.96 | 0.95 | 0.94 | 0.93 | 0.92 | 0.85 | 0.84 | 0.6 | 0.2 | 0 |
| 0.079 | -0.5 | 1 Year $1044m | 2 Year $787m | 2 Years $530 | 3 Years $329m | 5 Years $226m | 8 Years $165m | 12 Years $124m | 22 Years $94m | $72m | $0m | $5m | $51m | $68m | $70m |
| 0.159 | -1 | 1 Year $1044m | 2 Years $530m | 5 Year $226m | 12 Year $124m | $72m | $41m | $21m | $6m | $5m | $41m | $44m | $68m | $76m | $80m |
| 0.238 | -1.5 | 1 Year $1044m | 3 Years $329m | 12 Years $124m | $55m | $21m | $0m | $13m | $23m | $31m | $55m | $57m | $72m | $84m | $90m |
| 0.318 | -2 | 1 Year $1044m | 5 Years $226m | $72m | $21m | $5m | $21m | $31m | $38m | $44m | $61m | $63m | $76m | $92m | $100m |
| 0.397 | -2.5 | 1 Year $1044m | 8 Years $165m | $41m | $0m | $21m | $33m | $41m | $47m | $51m | $67m | $68m | $80m | $100m | $111m |
| 0.477 | -3 | 1 Year $1044m | 12 Years $124m | $21m | $13m | $31m | $41m | $48m | $53m | $57m | $69m | $69m | $84m | $109m | $121m |
| 0.556 | -3.5 | 1 Year $1044m | 22 Years $94m | $6m | $23m | $38m | $47m | $53m | $57m | $60m | $70m | $71m | $88m | $117m | $131m |
| 0.636 | -4 | 1 Year $1044m | $72m | $5m | $31m | $44m | $51m | $57m | $60m | $63m | $72m | $72m | $92m | $125m | $142m |
| 0.715 | -4.5 | 1 Year $1044m | $55m | $13m | $36m | $48m | $55m | $59m | $63m | $66m | $73m | $74m | $96m | $133m | $152m |
| 0.794 | -5 | 1 Year $1044m | $41m | $21m | $41m | $51m | $57m | $61m | $66m | $68m | $75m | $76m | $100m | $142m | $162m |
| 0.874 | -5.5 | 1 Year $1044m | $30m | $26m | $45m | $55m | $59m | $64m | $67m | $68m | $76m | $77m | $105m | $150m | $172m |
| 0.953 | -6 | 1 Year $1044m | $21m | $31m | $48m | $57m | $61m | $66m | $68m | $69m | $78m | $79m | $109m | $158m | $183m |

Key:
- Blue: Profitable and breaks even in less than 30 year
- Yellow: Profitable and breaks even in more than 30 years
- Red: Not profitable

X years: The number of years necessary to break even
$Ym: The profit per year

returns you receive from holding assets such as bonds, this increases the opportunity cost of holding precious metals such as platinum which do not yield any interest or returns. If the market expects with a high probability that the Federal Reserve will increase interest rates, then the demand from speculators for precious metals such as platinum will fall and with it the price of platinum.

Platinum is not a self-contained market unaffected by other markets and factors in the global economy. For example, the supply of platinum can be affected by movements in currency values. In general platinum miners and refiners pay their costs such as wages, equipment and electricity, in the local currency of the country in which they are mining but receive their revenue in foreign currency as their platinum is largely sold abroad as exports. This means that if the local currency decreases in value relative to other currencies that profit margins for the platinum miner increase, because the revenue of foreign currency they receive for their platinum increases in local currency terms, whilst their costs remain unchanged in local currency terms.

An example of this is occurring in South Africa, which over the period 2013-2017 produced 53% of the global supply of platinum [34], where we define the global supply as including mined and recycled platinum. Despite the platinum prices being at their lowest point in a decade, platinum producers in South Africa are not decreasing production, in fact their largest producer, Anglo American Platinum Ltd., is increasing its production [48]. This has partly been made possible due to the South African Rand being weak and so supporting the profit margins of South African producers.

The global demand for platinum can also be affected by the movement of currency values. For example, changes in the strength of the dollar can change the effective price Americans face for platinum in dollar term, and so the American demand for platinum. Since America constituted 11% of global demand for platinum on average in the period 2013-2017 [34], changes in the relative strength of the American dollar can affect global platinum demand and therefore prices.

There are also black swans that can affect the price of platinum. An example of this is how the Volkswagen scandal affected consumers opinions and government treatment of diesel cars, contributing to reducing diesel car sales, meaning reduced demand from diesel car producers for the platinum that is used in catalytic converters [49]. Given that in the years 2013-2017 the demand for platinum as an auto catalyst for diesel cars constituted 39% of the global demand for platinum [34], anything which affects this element of platinum demand will therefore strongly affect global platinum demand and therefore platinum prices.

The model indicates that a platinum asteroid miner would only be profitable in a relatively small and unlikely set of conditions for the global platinum market, this draws into question whether a platinum asteroid miner would be able to survive the volatility that can occur in the global platinum market.

## 5. Discussion

The results from the profitability model show that first and foremost, the economic viability of an asteroid mining venture depends on the throughput rate, which depends on the mining process. This raises interesting questions about the properties of the mining and refining process. Furthermore, development cost is an important factor and could be reduced via developing smaller spacecraft but using several of them per mission. In terrestrial industrial processes, scaling up a process usually leads to an increase in efficiency. In an asteroid mining context, this could mean that either the spacecraft is scaled up in size or the number of spacecraft is increased, in order to reduce cost by economy of scale effects. However, our results indicate that economy of scale effects for typical learning curve values do not make a significant difference for breakeven. Reuse can also help to reach profitability more quickly but needs to be applied with care, as expendable spacecraft may be more be advantageous due to a compressed sequencing of missions.

Regarding specific mining missions, our results have confirmed that mining volatiles is easier from a throughput rate point of view due to the higher fraction of volatiles in C-type asteroids. This is a supplement to the common argument that volatiles are simply easier to mine due to their lower boiling point compared to metals.

The supply-demand analysis has shown that supplying platinum from space is only economically viable in a very narrow set of values for price elasticity and substitution. For realistic price elasticity values, the mining venture would only be economically feasible when the quantity of platinum from space would substitute an equal quantity of terrestrial platinum.

## 6. Conclusions

In this article we analysed the economic feasibility of asteroid mining, focusing on supplying water in space and returning platinum to earth. We find that from a profitability perspective, the throughput rate and using smaller but multiple spacecraft per mission are key technical parameters for reaching breakeven quickly. Hence, the development of efficient mining processes and developing small spacecraft that are mass-produced are key to economic viability. In particular, platinum mining requires very high throughput rates, which could be difficult to achieve. Furthermore, for returning resources from space to Earth, the reaction of the Earth-based market is critical for economic viability. As in previous studies, mining volatiles and supplying them to cis-lunar orbit seems to be economically viable without the development of mining and refining processes with very high throughput rates.

For future work, we propose to explore the role of regulatory instruments for supplying resources from space to Earth. In addition, the estimation of possible throughput rates of known mining and refining processes would merit further investigation.